\documentclass[12pt,aps,prd,preprint,tightenlines,
   showpacs,nofootinbib]{revtex4}
\newcommand{\PRE}[1]{{#1}} 

\usepackage{bm} \usepackage{epsfig}
\newcommand{\eqref}[1]{Eq.~(\ref{#1})}

\newcommand{\secref}[1]{Sec.~\ref{sec:#1}}

\newcommand{\figref}[1]{Fig.~\ref{fig:#1}}

\begin{document}

\preprint{UCI-TR-2006-23}

\title{
\PRE{\vspace*{1.5in}}
Discovering SUSY with $m_0^2 < 0$ in the First LHC Physics Run

\PRE{\vspace*{0.3in}} }

\author{Arvind Rajaraman}
\affiliation{Department of Physics and Astronomy, University of
California, Irvine, CA 92697, USA \PRE{\vspace*{.5in}} }
\author{Bryan T.~Smith%
\PRE{\vspace*{.2in}} } \affiliation{Department of Physics and
Astronomy, University of California, Irvine, CA 92697, USA
\PRE{\vspace*{.5in}} }

\begin{abstract}
\PRE{\vspace*{.3in}}
In minimal supergravity, the parameter space
where the  slepton is the LSP is usually neglected, because of strong
constraints on charged dark matter. When the gravitino is the true LSP,
this  region avoids these constraints
and offers spectacular collider signals. We investigate this scenario
for the LHC and find that a large portion of the ignored mSugra
parameter space can lead to discovery within the first physics run,
with 1-4 $\text{fb}^{-1}$ of data. We  find that there are regions
where discovery is feasible with only 1 day of running.


\end{abstract}


\maketitle

\section{Introduction}
\label{sec:introduction}

Some of the most highly motivated extensions of the standard
model (SM) are models with supersymmetry (SUSY). 
SUSY solves the hierarchy problem, and provides a natural dark matter
candidate if R-parity is exact. Furthermore, supersymmetry will be
tested in a few years at the upcoming Large Hadron Collider (LHC).

 On
the other hand, supersymmetry is manifestly not a symmetry at low
energies, and hence supersymmetry must be broken. The most general
framework of supersymmetry breaking
already has over a hundred parameters, and any analysis requires a
simplified model. The most popular of these simplified models is
minimal
 supergravity
(mSugra)~\cite{Chamseddine:1982jx}, which is usually taken to be
specified by 5 parameters
$(m_0^2, M_{1/2}, A_0, \tan\beta, \ \text{and}\ \text{sign}(\mu))\ $,
which are respectively
the  scalar mass squared, the  gaugino mass, the trilinear term, the
ratio of the   up and down type Higgs boson vacuum expectation
values, and the supersymmetric Higgs mass parameter. The first three
are evaluated at the unification scale. Low energy parameters, like
superpartner masses and couplings, can be determined by the
renormalization group evolution of these parameters.

One of the constraints which is usually imposed on the parameters of
mSugra is that regions of parameter space where the lowest
superpartner is a slepton are taken to be ruled out. This is because
the lightest superpartner (LSP) is stable if R-parity is conserved
(which we will assume henceforth.) If the slepton is the LSP, it
appears as an absolutely stable charged massive particle (CHAMP), and
there are very strong bounds on CHAMP masses, both from direct
searches~\cite{Smith:1979rz,Dimopoulos:1989hk}, and from cosmological
effects on nucleosynthesis~\cite{Pospelov:2006sc,Cyburt:2006uv}. This
constraint removes a large region of mSugra parameter space. In the
$(m_0^2, M_{1/2})$ plane, all parameter values with $m_0^2 <0$ lead
to a slepton LSP, and are excluded by this constraint. A thin
triangular wedge in the $m_0^2 > 0$ parameter space is also excluded.

However, as pointed out in~\cite{Feng:2005ba}, this constraint does
not consider effects from the gravitino. In particular, if we are in
the parameter space where $m_0^2<0$, we can consider the possibility
that the gravitino is the true
LSP~\cite{Feng:2003xh,%
Ellis:2003dn,Feng:2004zu,Wang:2004ib,Ellis:2004bx,%
Roszkowski:2004jd,Brandenburg:2005he,Cerdeno:2005eu}. In that case,
the charged slepton will be the NLSP, and charged dark matter
particles will decay to the gravitino in a short time period. The
current dark matter will then be entirely composed of gravitinos, and
CHAMP searches will not put any constraints on this scenario.
Nucleosynthesis bounds are satisfied if the slepton lifetime is less
than about $10^3$ seconds~\cite{Feng:2003xh,Feng:2004zu,Roszkowski:2004jd,%
Cerdeno:2005eu,Jedamzik:2004er,Kawasaki:2004qu,Ellis:2005ii,Jedamzik:2005dh}.
(In fact there are many cosmological virtues of a gravitino LSP,
particularly in aspects regarding structure formation
~\cite{Kaplinghat:2005sy,Cembranos:2005us,Jedamzik:2005sx,%
Sigurdson:2003vy,Profumo:2004qt}.) One can thus find models in the
region with $m_0^2<0$ which are consistent with all current limits.

This region of parameter space has unique signatures. The NLSP
sleptons are stable on collider timescales, and exit the detector
before decaying.
Thus the signal of this region of SUSY parameter space at colliders
will be heavy muon-like particles.
This is radically different from the commonly considered situation
where the neutralino is the LSP. In that situation, the neutralino
escapes undetected, and supersymmetry in manifested in events with a
large missing energy. In this new region, the signatures are quite
different,
and standard missing energy searches for SUSY may { completely miss}
this entire region of parameter space.

In this paper, we will perform the first comprehensive analysis of
the slepton NLSP region, with the primary aim of finding out how soon
this region of  parameter space can be discovered at the LHC. In
particular, we will show that a large portion of parameter space can
be probed at the first high energy run of the LHC with no more than
1-4 fb$^{-1}$ of data, and some points may have a ``Day 1'' discovery
of SUSY.

While there has been a lot of earlier theoretical work on long lived
sleptons in supersymmetric theories, it has mostly been focused on
gauge-mediated SUSY breaking (GMSB) models (see eg.
~\cite{Hinchliffe:1998ys,Ambrosanio:2000ik}), and mSugra in the
region where $m_0^2>0$ (e.g.~\cite{Ellis:2006vu}).
 Furthermore, these papers
 were oriented toward an analysis of mass reconstruction, and hence focused
on special benchmark points. A complete scan of the slepton NLSP
region (along the lines of \cite{Baer:1995nq,Baer:1995va} for the
neutralino LSP region) has not yet been performed, and is one of the
goals of this paper.

There has also  been a lot of experimental work searching for
long-lived charged particles at colliders, again typically focused on
GMSB models. Experimental searches at LEP and the Tevatron have so
far given null results, putting constraints on the masses and cross
sections of these particles~\cite{Experiment:1995al}. There have also
been a number of experimental studies of the detection of these
particles at the LHC \cite{AtlasTDR}. We shall review the
experimental search strategies below in \secref{general} , and see
how they can be utilized to probe this new parameter region.


After a discussion of the experimental signals, we apply our analysis
to two benchmark points, which were previously discussed in
\cite{Feng:2005ba}.  We find a set of cuts to isolate the signal from
the background, and
show 
that both points can be discovered in the first physics run of the
LHC. In \secref{scan} we extend this to  a scan over the extended
mSugra parameter space (including the previously ignored $m_0^2<0$
region). We show that much of this parameter space
can be probed with 1-4 fb$^{-1}$ of data. 
We conclude with a discussion of our results in \secref{conclusion}.

\section{General Considerations}
\label{sec:general}

Since the slepton appears in the detector as a heavy muon, one of the
crucial elements in an analysis of the $m_0^2<0$ parameter space is
to find a way to distinguish these sleptons from the large background
of muons produced from Standard Model processes. There are two
important observables which are commonly used to perform this
separation: time-of-flight (TOF) measurements and ionization loss
(dE/dx) measurements.

\subsection{Time-of-flight (TOF)}The TOF method uses the fact that a slepton
is moving significantly slower than a muon of the same momentum. 
Every muon is travelling at essentially the speed of light.
The sleptons on the other hand have a velocity
$v=\beta_{\tilde{l}}c=p_{\tilde{l}}c/E_{\tilde{l}}$. Since the masses
of the sleptons are at least 100 GeV, the corresponding
$\beta_{\tilde{l}}$ can be significantly different from 1 even at LHC
energies.

The smaller velocity of the sleptons means that they arrive at the
farther parts of the detector (typically the muon chambers) at a
later time than the muons. The time delay is  $\Delta t={d\over \beta
c}-{d\over c}$ ($d$ is the size of the detector). By measuring this
time delay, the velocity can be calculated. By requiring the velocity
to be significantly different from the speed of light, we can remove
most of the muon background.

The resolving power of this method depends on both the size of the
detector and the time resolution. The time resolution of ATLAS and
CMS is the same at 1 ns. Since the ATLAS detector is bigger, it has a
better resolving power.
We will therefore focus on the ATLAS detector for the remainder of
this paper.


ATLAS's muon system measures the TOF through its resistive place
capacitors (RPC) for $\eta < 1$ and thin gap capacitors (TGC) for $1
< \eta < 2.4$.  For $\eta > 2.4$ a TOF measurement can not be made,
and hence we limit our study to sleptons candidates within $0 < \eta
< 2.4$.

\subsection{Ionization loss} While the TOF technique achieves a very clean separation between muons
and sleptons, it also leads to a huge decrease in the signal, since
most sleptons are produced with $\beta$ very close to 1. Accordingly
it is useful to find other methods to reduce the background.

The second technique to separate muons from sleptons is through the
measurement of ionization energy loss (sometimes referred to as the
measurement of dE/dx). The ionization energy loss of a charged
particle depends on $\beta$, and is given by the Bethe-Bloch formula.
By measuring the energy deposition as the charged particle passes
through the detector, one can estimate its velocity, and thereby
distinguish muons from sleptons. (For discussion of this method in
the context of R-hadrons, see \cite{Kraan:2005ji}).

The muon system is capable of measuring the ionization energy loss,
through the monitor drift tubes (MDT),
which measure the total charge ionized by a charged particle that
travels through them.  Slow moving sleptons will deposit more charge
in the MDTs than a muon.  It has been shown that this charge
difference can be measured and used to distinguish between long-lived
sleptons and muons~\cite{Polesello:1999aa}. In this paper, it is
claimed that the background can be reduced by $10^8$ while reducing
the signal only by a factor $10^2$.


In addition, the 
Transition Radiation Tracker 
can be used to distinguish ultra relativistic particles from slower
moving particles \cite{TRT}.  Charged particles passing through the
ATLAS TRT deposit ionization energy, and particles with
$\beta\gamma>1000$ also lose energy through transition radiation.
This energy can be picked up by the straws in the TRT. A particle
that deposits energy greater than 5.5 keV is recorded as a
high-threshold hit for that straw, while if the deposition is above
the lower threshold of 200 eV, the time for which the signal is above
the threshold is also recorded. Both the number of high-threshold
hits and the total time over-threshold for the low threshold hits
depend on the
 total energy deposition.
 We can therefore use the TRT to separate slow particles from ultrarelativistic
 particles.

A combined study using all these signals has apparently not yet been
performed. Using the ionization energy loss and transition radiation
measurements would therefore require a full detector simulation,
which is in progress~\cite{UCI}. 
 For this paper, we will not consider ionization loss, and instead
focus on the TOF measurements and appropriate kinematic cuts to
reduce  the background from SM muons.

\section{Analysis of Benchmark Points}
\label{sec:benchmark}

\subsection{Spectra and Signatures}

\begin{table}
\begin{tabular}{|c|c|c|c|c|c|}
\hline
&$m_0^2$&$M_{1/2}$&$A_0$& $\tan\beta$&sign($\mu$) \\
\hline Model A& $-(40)^2\ {\rm{GeV}}^2$&\ $300\ {\rm{GeV}}$ &\ $0\
{\rm{GeV}}$ &\ 10 &\ +\\
\hline Model B& $-(700)^2\ {\rm{GeV}}^2$&\ $1900\ {\rm{GeV}}$ &\ $0\
{\rm{GeV}}$ &\ 60 &\ +\\
\hline
\end{tabular}
 \caption{Parameters of our two benchmark models.}
 \end{table}

\begin{figure}
\resizebox{6.0in}{!}{
\includegraphics{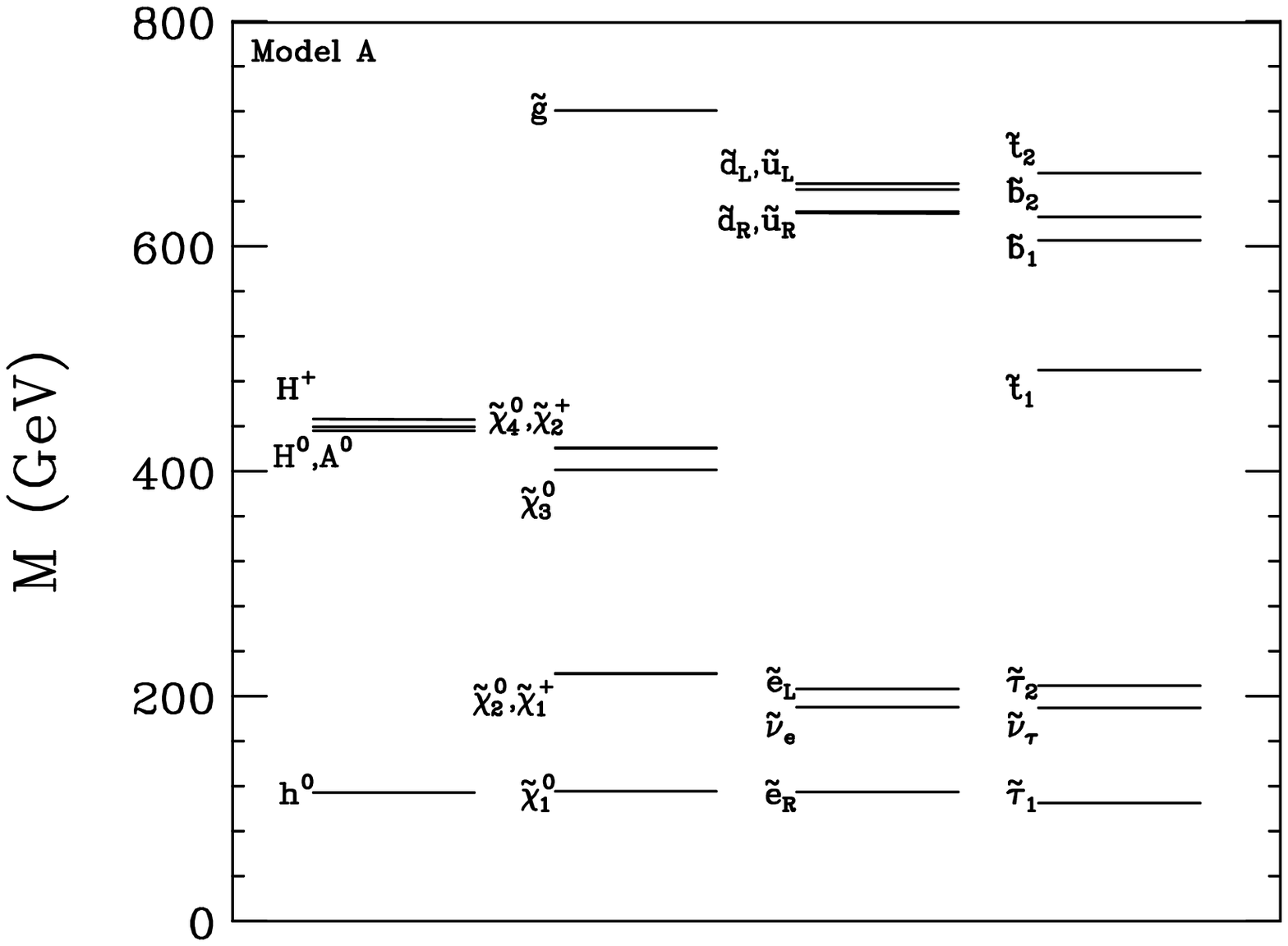} \qquad
\includegraphics{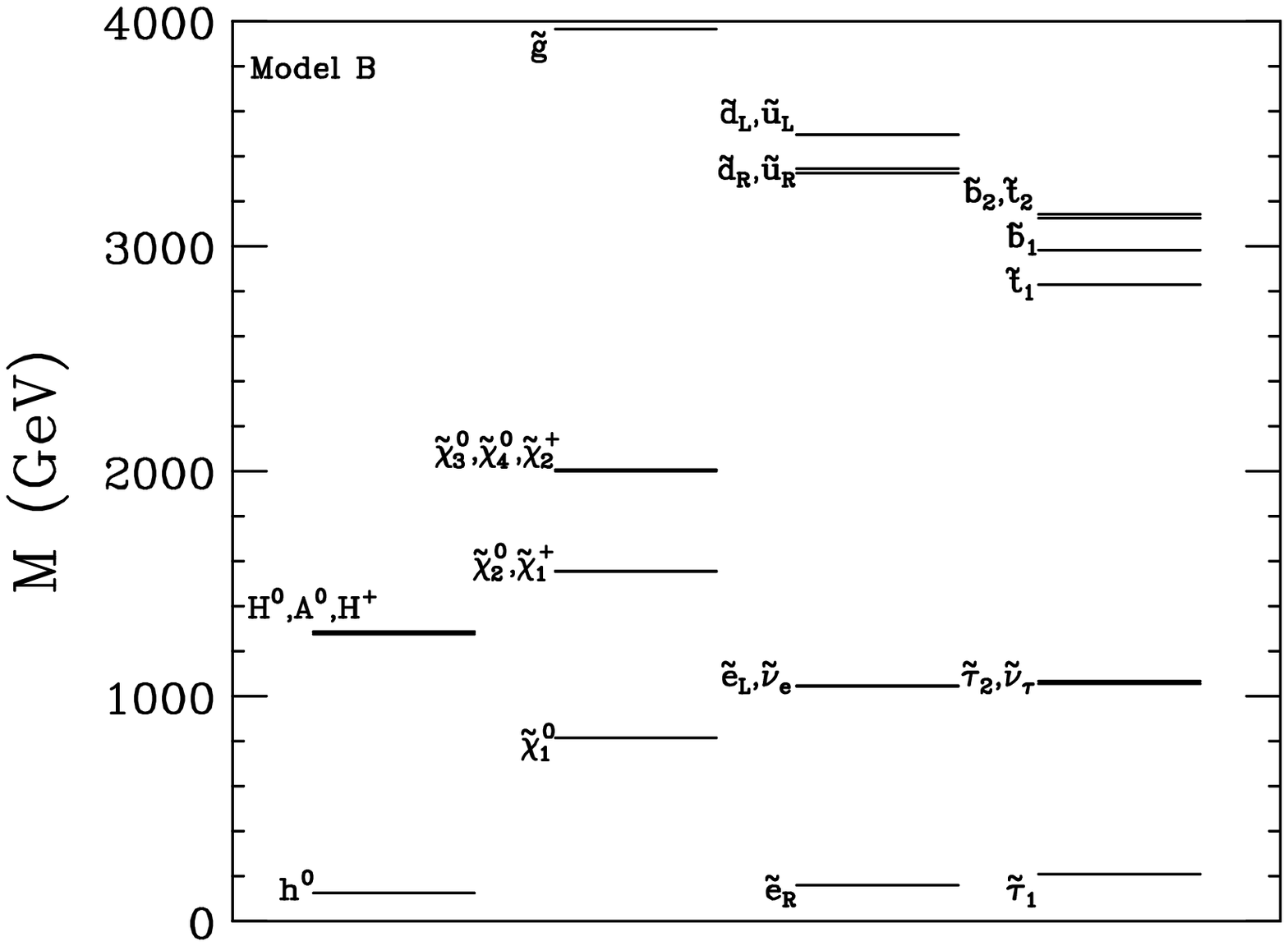}
} \caption{The SUSY mass spectrum for model A~(left) and model
  B~(right).  The first column shows the Higgs bosons, the second column
  shows the gauginos, the third column shows the first
  generation scalar particles, and the fourth column shows the third
  generation scalar particles.\label{fig:BenchMass}}
\end{figure}

We start by looking at two benchmark models, labelled model A and
model B, from our previous paper~\cite{Feng:2005ba}.  The mass
spectra of the two models are shown in~\figref{BenchMass} and are
produced from ISAJET v7.71~\cite{Paige:2003mg} modified to calculate
the SUSY mass spectrum for $m_0^2 < 0$.

Each of the models has a different signature. Model A is a ``typical''
SUSY spectrum where squarks and gluinos dominate the production of
SUSY particles.  The colored SUSY particles cascade down to the NLSP
charged lepton.
The signature is two hard jets, two charged staus, and two leptons.
In this model the cascade decay of the squarks and gluinos produces a
large amount of transverse energy.

 Model B, on the
other hand, has a mass spectrum where Drell-Yan production of the
light charged sleptons dominates the SUSY production.  There are  two
charged sleptons in the final state, and no jets.


In addition, the velocity distributions of the sleptons are somewhat
different. More slow sleptons are produced in Drell-Yan processes as
compared to cascade decays.
The TOF method therefore works better for the situations where
Drell-Yan processes dominate, as in model B.

\subsection{Background analysis}
The  background comes from  Standard Model  events in which two muons
are produced. The dominant sources of these events are listed in
table. \ref{table_background}.

We generated $10^6$ events of each source with PYTHIA
6.404~\cite{Sjostrand:2006za}. To reduce these backgrounds we use a
series of cuts:

{\it a. Muon number:} Two muon-like particles must be produced.

{\it b. Rapidity cuts:} Both muon-like particles must be within
$\eta<2.4$.

 {\it c. Isolation
cuts:} Both muon-like particles should contain less than 10 GeV of
transverse momentum in a cone of
$R=\sqrt{(\Delta\eta)^2+(\Delta\phi)^2}<0.2$ around each muon-like
particle.

 {\it d. Momentum
cuts.} Both muon-like particles must have a momentum greater than 100
GeV and a transverse momentum greater than 20 GeV.

\vskip 0.3cm

Once these cuts are imposed, the dominant background to Drell-Yan
production is Z-production. To reduce this background, we add a cut
on the invariant mass. We thus define the

 {\it e. Drell-Yan cuts}: The event should pass cuts (a), (b), (c), (d);
 in addition,  the invariant mass of the two muon-like
 particles must be greater than 120 GeV.

\vskip 0.3cm

For the cascade decays,  on the other hand, the dominant background
after the cuts (a), (b),(c), (d)  is Z+jets. To reduce this
background we impose a requirement of having hard jets. We therefore
define the

 {\it  f. Cascade Cuts}: The event should pass cuts (a), (b), (c), (d);
 in addition, we require that the 4 most energetic objects among the jets
and leptons each have an energy greater than 70 GeV.

\vskip 0.3cm

 \begin{table}
\begin{tabular}{|c|c|c|c|c|}
\hline
&Total cross-section& After Drell-Yan cuts&  After Cascade cuts\\
\hline
Model A&\ $18$pb &\ $9$pb &\ 8pb\\
Model B&\ $43$fb &\ $28$fb &\ 1fb\\
\hline
QCD&$10^{2}$mb&$<1$pb &\ $<1$pb\\
$\gamma^*/Z\rightarrow \mu\mu$&100nb&3pb &\ 100fb\\
W+jet&360nb&$<40$fb &\ $<40$fb\\
Z+jet&150nb&7pb &\ 300fb\\
$t\bar{t}$&800pb&430fb &\ 80fb\\
WW,WZ,ZZ&$2.5$nb&150fb &\ 25fb\\
\hline
\end{tabular}
 \caption{Signal and backgrounds for Drell-Yan and cascade cuts
 \label{table_background}}
 \end{table}

The results of these cuts on the signal and background can be seen in
Table~\ref{table_background}. To estimate the QCD background we
generated $10^6$ events with
 $b\bar{b}$. We found no events passing these cuts, which
 corresponds to a direct limit on this cross-section of 700 fb. We
 expect the other QCD processes to contribute even fewer events, and
 so we have very conservatively estimated the cross section from QCD processes to be less
 than 1 pb.

\subsection{TOF cuts}

Finally, we can reduce the background by using the time-of flight
methods as described in section \ref{sec:general}. The ionization
loss method could also potentially be used, but we shall not use it
in this analysis.

\begin{figure} \resizebox{6.0in}{!}{
\includegraphics{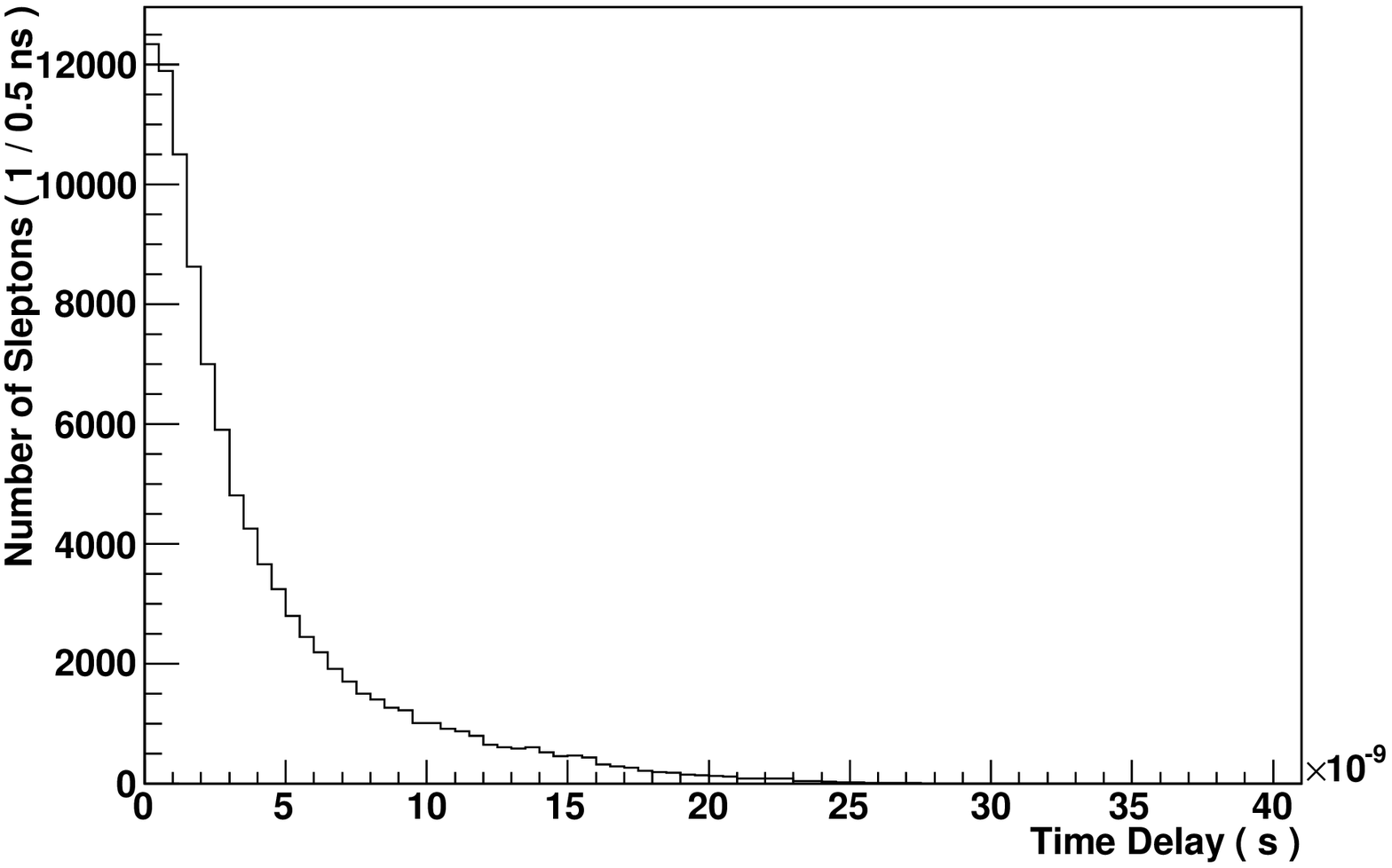} \qquad
\includegraphics{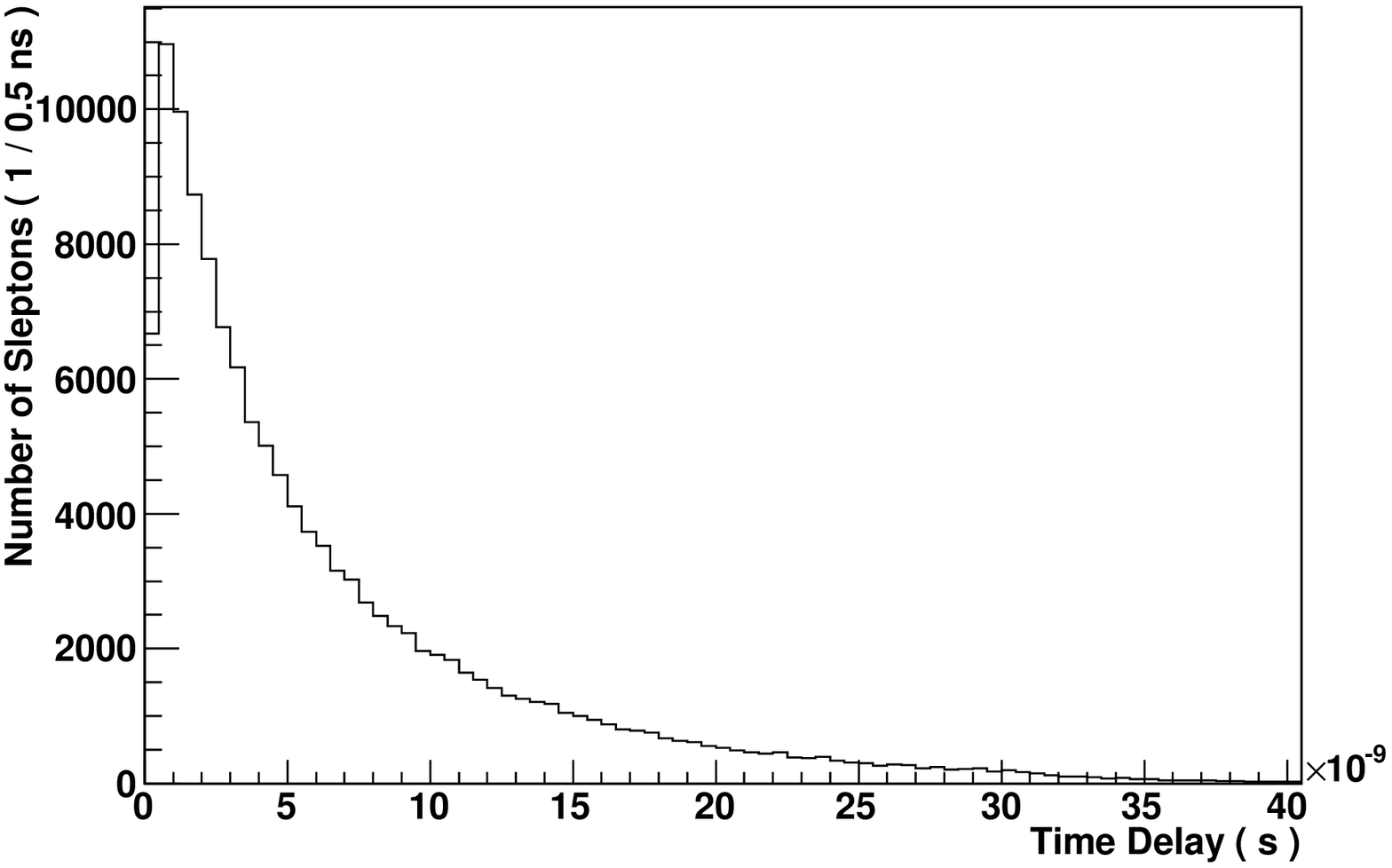}
} \caption{The distribution of the time delay for $2 \times 10^5$
sleptons that
  survive the Drell-Yan cuts. 
  The left panel shows results for model A 
  and the right panel shows results for model B.\label{fig:timedelay}}
\end{figure}

The background SM muons are required to have a momentum larger than
100 GeV by our cuts.  The time delay of such muons to the farthest
part of the detector is orders of magnitude smaller than  the
detector resolution of 1 ns. (The sleptons, on the other hand, have a
significant time delay
as seen in \figref{timedelay}.) However, mismeasurements of the time
can lead to apparent time delays. The probability of a mismeasurement
of
$\Delta t$ ns, is expected to go as
exp(-${\Delta t}^2$),
assuming a gaussian resolution to the TOF measurement.
We will use a TOF cut where we require {\it both} muons to have a
time delay greater than some value ${\Delta t}$. Our TOF cut
therefore reduces the SM background by a factor of exp($2{\Delta
t}^2$).

In Table~\ref{timedelay_table} the signal and background of our
benchmark points are shown for TOF cuts between 1-5 ns.
For a 3 ns time delay cut on each slepton we reduce our background to
less than 1ab, which is  negligible (there will be no background
events over the entire running of the LHC.)

 \begin{table}
\begin{tabular}{|c|c|c|c|c|c|c|}
\hline
Time delay of&0ns&1 ns& 2ns &3ns&4ns&5ns\\
\hline Drell-Yan; background&\ ~10pb &\ ~1.35pb &\ ~3.3fb &\ ~0.2ab
&\
$<0.1$ab &\ $<0.1$ab\\
Drell-Yan; Model A&\ 9pb &\ 5.2pb &\ 2.9pb &\ 1.8pb &\ 1.1 pb &\ 750fb\\
Drell-Yan; Model B&\ 28fb &\ 23fb &\ 18 fb&\ 14fb &\ 11 fb&\ 9.4fb\\
\hline Cascade; background&\ $<1$pb &\ $<134$fb &\ $<340$ab &\ $<0.1$ab&\ $<0.1$ab&\ $<0.1$ab\\
Cascade; Model A&\ 8pb &\ 4.3pb&\ 2.4pb&\ 1.4pb&\ 910fb&\ 590fb\\
Cascade; Model B&\ 190ab &\ 87ab &\ 41ab &\ 22ab &\ 13ab &\ 7ab \\
\hline
\end{tabular}
 \caption{Signal and backgrounds   as a function of TOF cut. \label{timedelay_table}}
 \end{table}

A discovery claim requires  $S > 10$ and $S/\sqrt{B} > 5$, where $S$ is
the number of signal events and $B$ is the number of background events.
The 3 ns time delay cut lowers the background so much that the second
requirement is automatically met if there are 10 signal events. The
only criterion for discovery is then the requirement $S > 10$.

With model A having a cross section after our Drell-Yan cuts of
$\sigma_{\text{A}} = 1.8\text{pb}$, an integrated luminosity of
$5.6\text{pb}^{-1}$ is required for discovery. For model B with a
cross section of $\sigma_{\text{B}} = 14 \text{fb}$, an integrated
luminosity of $720 \text{pb}^{-1}$ is required for discovery.  Both
benchmark points can be discovered in the first physics run at the
LHC, where the expected luminosity is between 1-4 fb$^{-1}$. Model A
could feasibly be seen on the very first day of the first physics
run.

\section{Scanning The Parameter Space}
\label{sec:scan}

\begin{figure}
\resizebox{6.0in}{!}{
\includegraphics{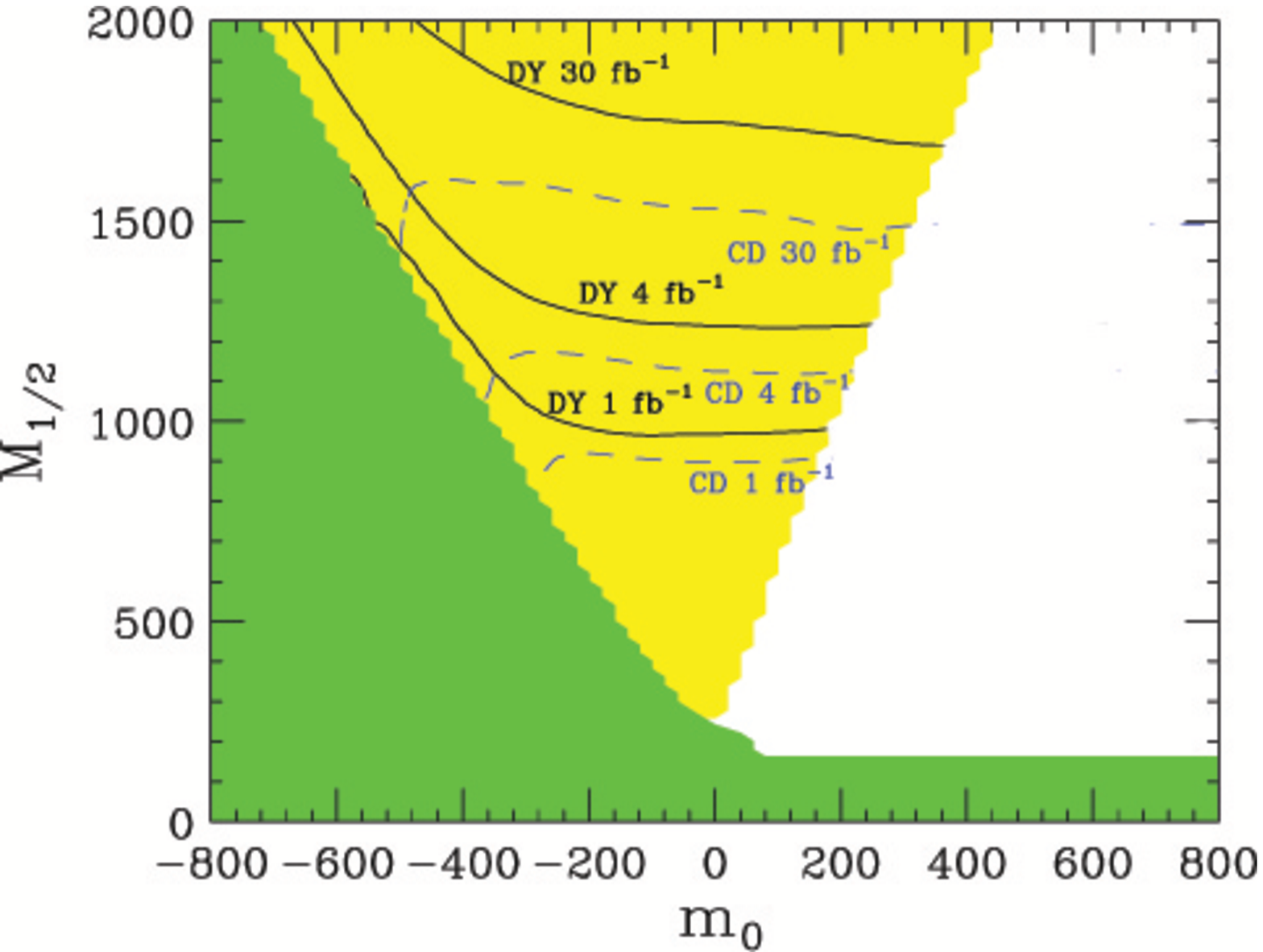} \qquad
\includegraphics{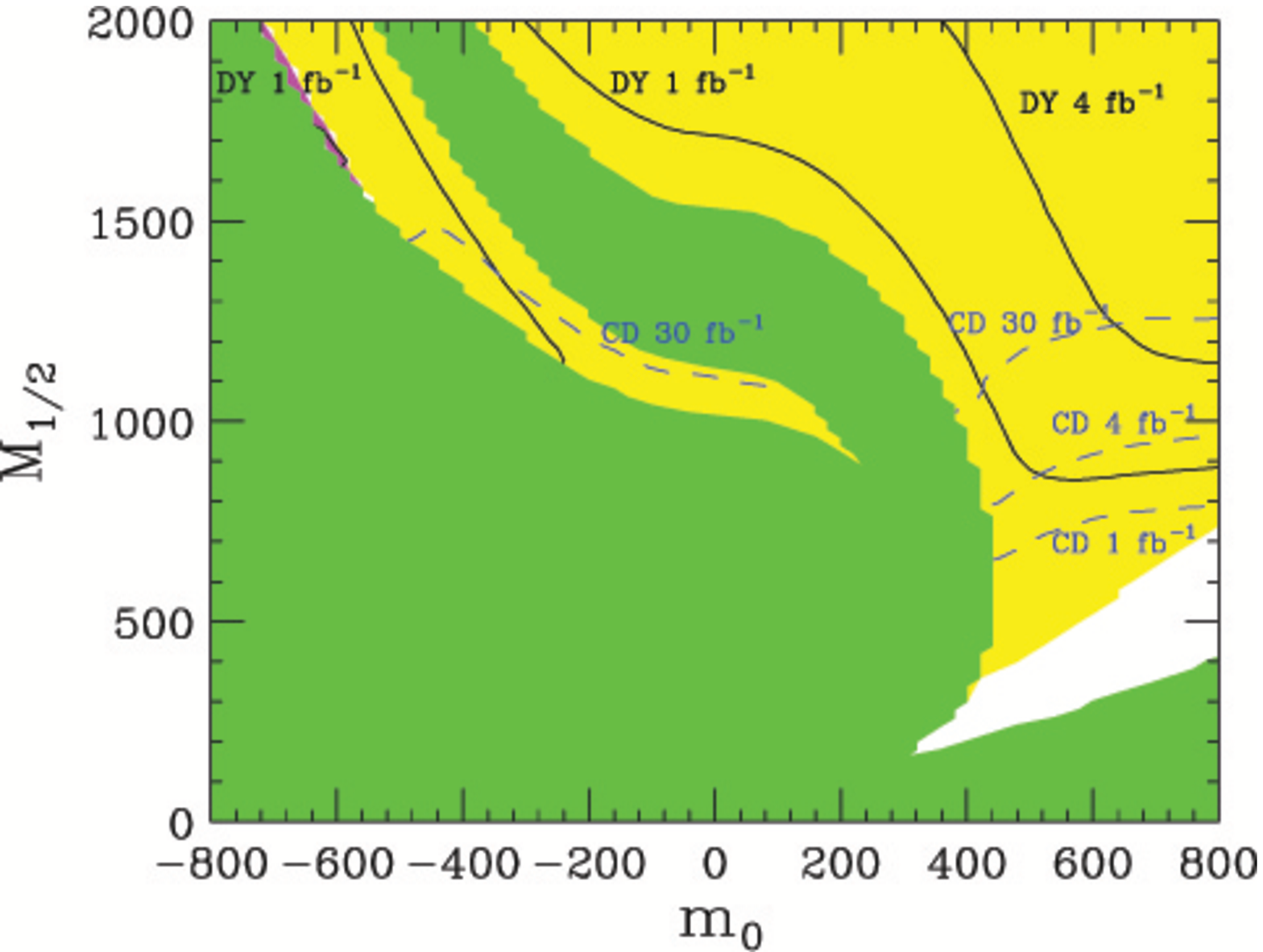}}
\caption{Luminosities required for discovery using either Drell-Yan
(DY) or cascade (CD) cuts.  Left panel is tan$\beta$~=~10, right panel is tan$\beta$~=~60 \label{fig:SUSY}}
\end{figure}

We now extend the analysis to a scan of the $(m_0^2,M_{1/2})$
parameter space with a slepton NLSP and gravitino LSP.  In the scan,
we set $A_0 = 0$ and $\mu > 0$ with tan$\beta$~=~10 and
tan$\beta$~=~60.    The contour plots in~\figref{SUSY} use the
notation $m_0 \equiv \text{sign}(m^2_0)\sqrt{|m^2_0|}$ and use the
following color scheme: green~(dark) for the experimentally excluded
region, yellow~(medium) for the stau LSP region, magenta~(darkest)
for the selectron LSP region, and white for the neutralino LSP
region.

 The mesh size  for our scan was 25 GeV for both $m_0$ and
$M_{1/2}$. For each point we generated 10000 events with PYTHIA 6.404
to get the fraction of events that pass the 3 ns time delay cut with
the Drell-Yan or cascade cuts, and the total SUSY cross section.

\begin{figure}
\resizebox{6.0in}{!}{
\includegraphics{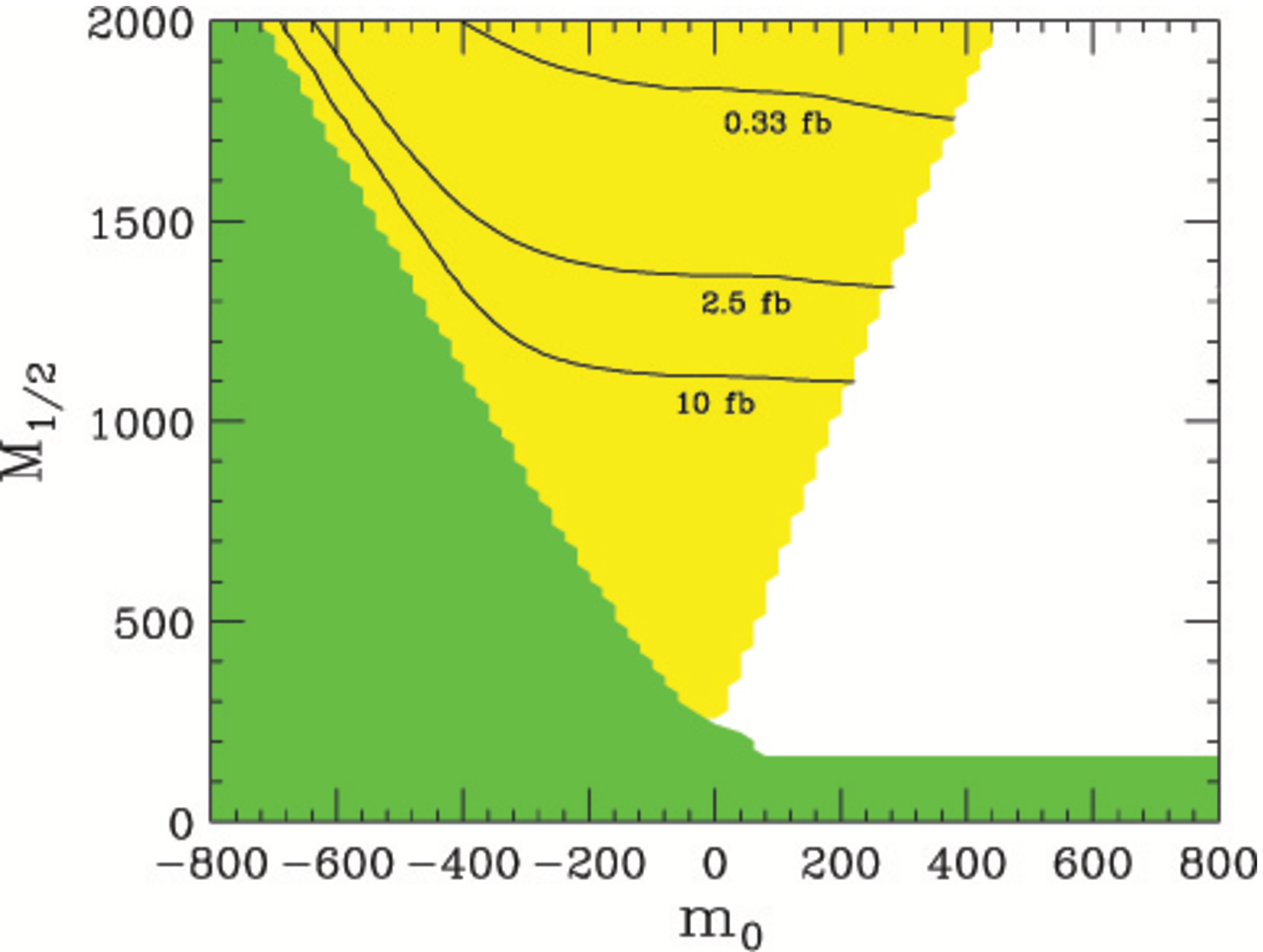} \qquad
\includegraphics{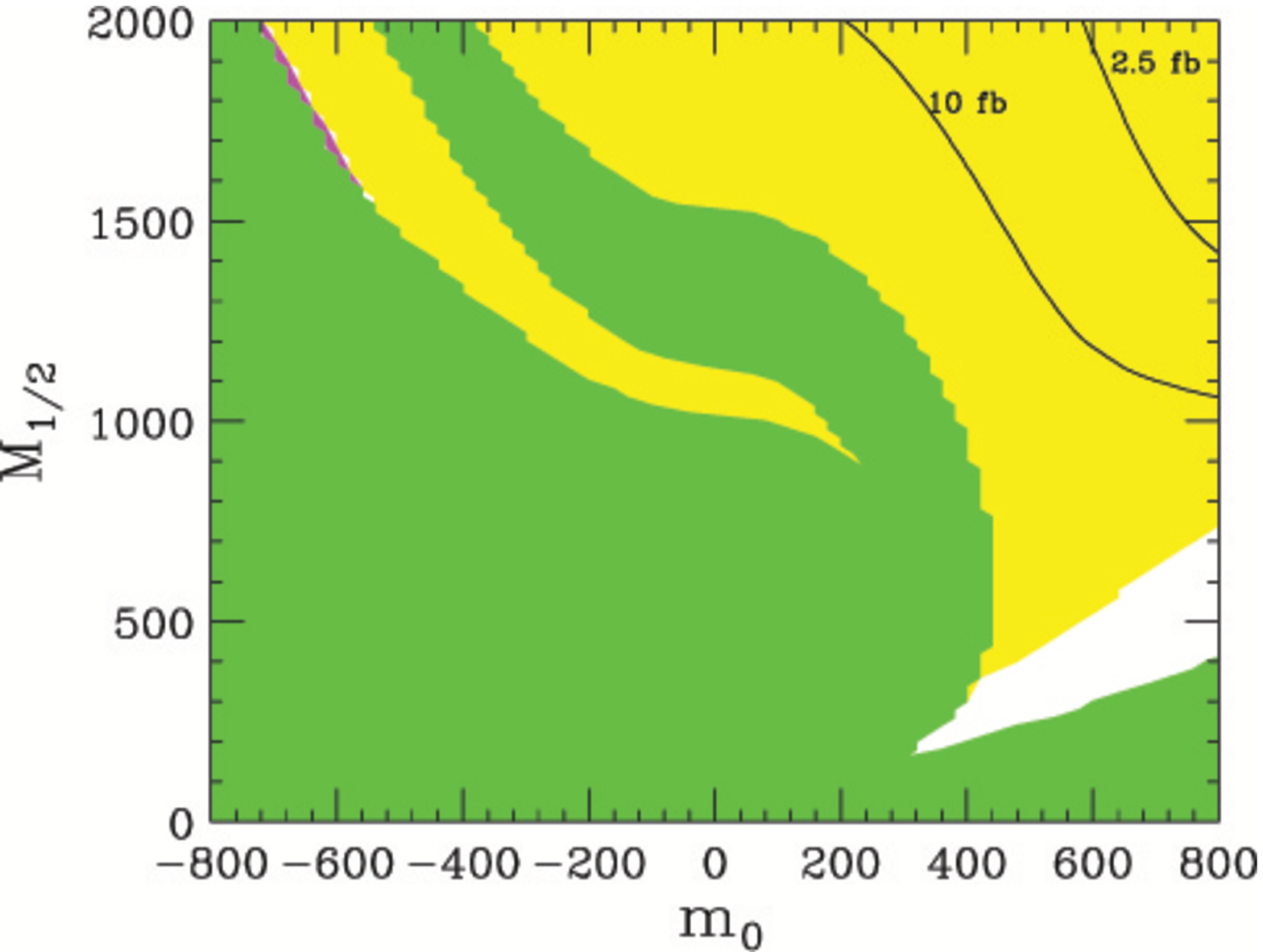}}
\caption{Total SUSY cross-sections prior to any cuts. Left panel is
tan$\beta$~=~10, right panel is tan$\beta$~=~60.
\label{fig:SUSYTOTAL}}
\end{figure}

The shape of the total SUSY cross sections, shown
in~\figref{SUSYTOTAL}, is determined by the mass
of the squarks and right handed sleptons. For larger $m_0$, 
the SUSY cross section tends to be dominated by squark and gluino
production. The SUSY cross section contours follow contours of
constant squark and gluino masses, which are flat as a function of
$m_0$. 
As $m_0$ decreases, the sleptons become lighter, and the SUSY
production is eventually  dominated by Drell-Yan production of right
handed sleptons. After this point the total SUSY cross section
follows contours of constant slepton mass, which are concave upward.


This can be checked by comparing the cross sections relative to the
mass contours plots in the extended mSugra region presented
in~\cite{Feng:2005ba}. For example, in the tan$\beta$~=~10 panels the
total SUSY cross section of 10 fb cross section starts on the right
following an approximately 2 TeV mass average of the squarks and
gluinos. As we follow the contour to the left, it turns up to follow
the contours of constant slepton mass of approximately 200 GeV.  A
similar behavior can be seen in the tan$\beta$~=~60 panels as well.

In \figref{SUSY}, the left panels show our results for
tan$\beta$~=~10, and 
the right panels show our results for for tan$\beta$~=~60. The total
SUSY cross section after time delay cut of 3ns and
 our Drell-Yan cuts~(solid) and cascade cuts~(dashed) is shown.
The cross sections are shown for three values: 10 fb, 2.5 fb, and
0.33 fb.
The 10 fb dashed lines correspond to 1 $\text{fb}^{-1}$ of integrated
luminosity for discovery at the LHC,  while the 2.5 fb and 0.33 fb
dashed lines correspond to discovery at 4 $\text{fb}^{-1}$ and
30 $\text{fb}^{-1}$ of integrated luminosity respectively.


One important (and somewhat surprising) result from this analysis is
that the Drell-Yan cuts have a better reach that than the cascade
cuts, even in regions dominated by cascade production. The basic
reason is that the Drell-Yan cuts remove less of the signal than the
cascade cuts.
It is true that the cascade cuts remove more SM background than the
Drell-Yan cuts, but since the background is negligible after TOF
cuts, this
does not lead to an improvement for the $S/\sqrt{{B}}$ ratio.



Our main result is that a significant portion of the extended mSugra
parameter space will be probed with our cuts in the first physics run
at the LHC, expected to have an integrated luminosity of
1-4 $\text{fb}^{-1}$.  A 4 $\text{fb}^{-1}$ integrated luminosity will
produce a discovery of long lived sleptons for all parameter space
with $M_{1/2} < 1000$ GeV as well as a large portion of  $M_{1/2} >
1000$ GeV for the more negative values of $m_0$.  All of the
presented parameter space with $m_0< 0$ for tan$\beta$~=~60 will be
probed with 4 $\text{fb}^{-1}$ integrated luminosity, while all of the
tan$\beta$~=~60 parameter space shown will be probed with
30 $\text{fb}^{-1}$.


\section{Conclusion}
\label{sec:conclusion}

The first physics run of the LHC is expected to start in 2008. The
first task of this run will be to calibrate and understand the
detectors.
The SM must also be understood at these new energies of 14 TeV
~\cite{Green:2006fa}.
An exciting prospect is that new physics can be discovered even with
just this first $\text{fb}^{-1}$ of data.

In this paper, we have analyzed a region of supersymmetric parameter
space where the slepton appears as a nearly stable particle. We have
shown that
it is 
possible that a discovery of such models  could be made with the data
from the first physics run.  Both of our benchmark models can be
discovered with
 the first $\text{fb}^{-1}$ of data.
In fact, 
a large portion of our extended mSugra parameter space (including the
$m_0^2 < 0$ region) can be probed in the first physics run of the
LHC.

Once the existence of new physics has been established, the nature of
this physics can be studied. For example, our benchmark model A
produces signals both from Drell-Yan production as well as from
cascade decays. By combining these signals, it may be possible to
calculate
both the squark and gluino masses as well as the masses of the right
handed sleptons. With sufficient statistics, it may also be possible
to calculate the spins of these particles, and thereby distinguish
SUSY scenarios from extra dimensional models. An analysis of this
will be presented in future work.

The discovery of  apparently stable charged particles at the LHC will
also have implications for cosmology. A stable charged particle would
be highly problematic for cosmology, and so it is likely that the
slepton would be metastable with a lifetime long on collider scales,
but short on cosmological scales (in the range $10^{-6}$ to $10^3$
seconds).
The lifetime of the long lived sleptons will need to be measured,
possibly by building traps as proposed
in~\cite{Hamaguchi:2004df,Feng:2004yi}.  The measurement of the
lifetime can give both the mass of the gravitino, which we assumed
the sleptons decays to in this extended mSugra framework, and a high
energy measurement of the Planck mass once the mass of the NLSP
slepton is constrained~\cite{Feng:2004gn}. This will also allow an
accurate prediction of the relic density, which can be compared to
cosmological data.

To conclude,
the regions of MSSM parameter space with slepton NLSP and gravitino
LSP will manifest themselves almost immediately in signals of heavy
muon-like objects.
Thus there are very exciting prospects of discovering supersymmetry
almost immediately at the LHC. The discovery of supersymmetry may be
just around the corner.

\begin{acknowledgments}
The authors would like to thank Mario Bondioli for helping us learn
and setup PYTHIA.  We would like to thank Jonathan Feng for useful
comments.   We would also like to
thank the participants of the UCI LHC Lunch meetings for valuable
discussion throughout this work. The work of AR is supported in part by
NSF Grant No.~PHY--0354993. BTS is supported in part by NSF CAREER
grant No.~PHY--0239817. 
\end{acknowledgments}


\end{document}